\documentclass[11pt]{article}
\pdfoutput=1

 \usepackage{jcappub}
\usepackage{caption}
\usepackage{hyperref}
\usepackage{graphicx}
\usepackage{amsmath}
\usepackage{mathtools}
\usepackage{physics}
\usepackage{siunitx}
\usepackage{tablefootnote}
\usepackage{soul}
\usepackage{color}
\usepackage{verbatim}
\usepackage{float}
\usepackage{bm}
\usepackage{ulem}
\usepackage[dvipsnames]{xcolor}
\usepackage{tikz}
\usepackage{orcidlink}
\usepackage[nameinlink, capitalise]{cleveref}
\usetikzlibrary{shapes.geometric, backgrounds, arrows, calc, fit}
\tikzstyle{process} = [rectangle, rounded corners, minimum width=8cm, minimum height=1cm, text centered, draw=black, fill=blue!30]
\tikzstyle{arrow} = [thick,->,>=stealth]
\usepackage{underscore}

\usepackage{color,appendix,latexsym,amsmath,amssymb,graphicx,booktabs,epsfig,hyperref,url}

\numberwithin{equation}{section}
\usepackage[english]{babel}

\usepackage{letltxmacro}
\LetLtxMacro{\originaleqref}{\eqref}
\usepackage[sorting=none]{biblatex} 
\addto\extrasenglish{%
}

\definecolor{MyBlue}{rgb}{0.15,0.15,0.70}
\hypersetup{
colorlinks=true,
citecolor=MyBlue,
linkcolor=MyBlue,
urlcolor=MyBlue
}

\definecolor{orange}{rgb}{0.98, 0.6, 0.01}
\definecolor{darkolivegreen}{rgb}{0.33, 0.42, 0.18}
\definecolor{tealblue}{rgb}{0.21, 0.46, 0.53}

\usepackage{xspace}

\newcommand{\Fone}{\ensuremath{W_{\mu}}}
\newcommand{\Ftwo}{\ensuremath{W^{\rm (NG2)}_\mu}}

\newcommand{\fpbh}{\ensuremath{f_{\rm PBH}}}
\newcommand{\fnl}{\ensuremath{\tilde{f}_{\rm NL}}}
\newcommand{\gnl}{\ensuremath{\tilde{g}_{\rm NL}}}
\newcommand{\betaz}{\ensuremath{\beta_{1-\rm PBH}}}
\newcommand{\calR}{\ensuremath{{\cal R}}}
\newcommand{\calRG}{\ensuremath{{\cal R}_{\rm G}}}
\newcommand{\sG}{\ensuremath{\sigma_{\rm G}}}
\newcommand{\erfc}{\ensuremath{{\rm ErfC}}}
\newcommand{\inverfc}{\ensuremath{{\rm InvErfC}}}

\usepackage{listings}
\definecolor{codegreen}{rgb}{0,0.6,0}
\definecolor{codegray}{rgb}{0.5,0.5,0.5}
\definecolor{codepurple}{rgb}{0.58,0,0.82}
\definecolor{backcolour}{rgb}{0.95,0.95,0.92}
\graphicspath{figures/} 

\title{Robust $\mu$-distortion constraints on primordial supermassive black holes from non-Gaussian perturbations}

\author[1]{Christian T.~Byrnes\,\orcidlink{0000-0003-2583-6536},}
\author[2]{Julien Lesgourgues \,\orcidlink{0000-0001-7627-353X},}
\author[2]{Devanshu Sharma\,\orcidlink{0009-0002-3302-2153}}

\affiliation[1]{ Department of Physics and Astronomy, University of Sussex, Brighton BN1 9QH, UK\\}
\affiliation[2]{Institute for Theoretical Particle Physics and Cosmology (TTK), RWTH Aachen University, \\ D-52056 Aachen, Germany}

\emailAdd{C.Byrnes@sussex.ac.uk}
\emailAdd{lesgourg@physik.rwth-aachen.de}
\emailAdd{drsharma@physik.rwth-aachen.de}

\date{}

\abstract{Explaining the origin of supermassive black holes via a primordial origin is severely challenged by the tight spectral distortion constraints on the amplitude of the primordial perturbations. Following the first calculation of how the $\mu$ constraints are modified by non-Gaussianity in a companion paper, we here make the first robust constraints on primordial black hole formation under large non-Gaussianity. Even the infinite $f_{\rm NL}$ limit is insufficiently non-Gaussian but much higher-order non-Gaussianity of the form $\calR=\calRG^5$ may allow the formation of any mass primordial black hole without conflicting with distortion constraints. We caution that such extreme models face other challenges.
}

\begin{document}

\begin{flushleft}
TTK-23-16
\end{flushleft}

\maketitle
\flushbottom

\section{Introduction}

Supermassive black holes (SMBHs) are observed in the centres of virtually all galaxies where good observations have been made, even at high redshift. Explaining the origin of these SMBHs remains a challenge by all proposed methods, such as the direct collapse of large gas clouds, runaway mergers, or massive accretion onto the first star remnants. Given that it is hard to find an astrophysical origin, many have speculated they could have a primordial origin. See \cite{Volonteri:2021sfo} for a recent review.

However, explaining SMBHs via a primordial black hole (PBH) seed is also challenging, primarily because of tight cosmic $\mu$-distortion constraints which limit the allowed amplitude of the primordial power spectrum on scales larger than about $10$ parsec, which includes the length scales relevant for SMBH formation if they are primordial \cite{Chluba:2012we,Chluba:2015bqa}. In general, the dissipation of acoustic waves in the baryon-photon plasma generated by density perturbations leads to spectral distortions which constrains the primordial power spectrum \cite{Sunyaev:1970er,Chluba:2012gq,Khatri:2012rt,Pajer:2012qep,Chluba:2013dna}. Assuming the initial density perturbations are Gaussian distributed the formation of even just one PBH with initial mass $\gtrsim 10^4 M_\odot$ is completely excluded by the $\mu$-distortion constraints. We note that the possibility of a PBH seed with mass around $10^3 M_\odot$ has also been considered and this is safe from spectral distortion constraints \cite{Khlopov:2004sc,Kohri:2014lza,Bernal:2017nec}, but one then still has to explain how accretion is sufficient for them to reach a much greater mass, already at high redshift.

The two primary routes (other than massive amounts of accretion) to evade the $\mu$-distortion constraints are either to invoke a PBH formation mechanism different from the standard mechanism of a direct collapse of large amplitude perturbations after horizon entry (for recent PBH reviews see \cite{Carr:2020gox,Green:2020jor,Escriva:2022duf,Green:2024bam}) or to invoke large non-Gaussianity \cite{Nakama:2017xvq,Unal:2020mts,Gow:2022jfb,Hooper:2023nnl}. In this paper we focus on non-Gaussianity, which is expected to help because PBHs are necessarily very rare objects which form deep in the tail of the probability density function (pdf), whilst spectral distortions are primarily generated by the most likely perturbations around the peak of the pdf. Therefore, it may be possible to reduce the average amplitude of the perturbations whilst adding large positive skewness to enhance the tail to generate PBHs. However, the assumption that the $\mu$ constraints won't also tighten has never been tested with an actual calculation -- with only the appendix of \cite{Juan:2022mir} making (an approximate) estimate of the impact of non-Gaussianity on the $\mu$-distortions\footnote{We note that several references have considered the impact of non-Gaussianity on spatial variations of the $\mu$-distortion \cite{Pajer:2012vz,Ganc:2012ae,Khatri:2015tla,Chluba:2016aln,Ravenni:2017lgw,Ozsoy:2021qrg,Zegeye:2021yml,Bianchini:2022dqh,Rotti:2022lvy,CMB-S4:2023zem}, but not on the global value. Combining these measurements may help to break the degeneracy between the variance and non-Gaussianity of the primordial perturbations.}. 

Because only a significant amount of non-Gaussianity is expected to successfully evade the $\mu$-distortion constraints, see e.g.~\cite{Nakama:2017xvq,Unal:2020mts,Gow:2022jfb,Hooper:2023nnl}, one should not expect the $\mu$ constraints to remain unchanged relative to the Gaussian $\mu$ constraints. In our companion paper -- Sharma et al \cite{paper1} -- we have made the first accurate calculation 
of the sky-averaged $\mu$-distortion subject to local non-Gaussianity, assuming that this non-Gaussianity is confined to the small scales that are relevant for PBH generation. This calculation covers any possible value of $\fnl$. Here, we see how those constraints combine with PBH constraints to test whether SMBHs could have a primordial origin. 

In this paper we show that local non-Gaussianity with primordial curvature perturbations obeying ${\cal R}={\cal R}_{\rm G}+3 f_{\rm NL} ( {\cal R}_{\rm G}^2- \langle{\cal R}_{\rm G}^2\rangle)/5$ is insufficient to change the bound on the allowed PBH mass in any significant way, even in the infinite $f_{\rm NL}$ limit, and that much higher-order non-Gaussianity is needed to evade spectral distortion constraints.

Electron-positron annihilation takes place while the horizon mass is around $10^6 M_\odot$. The consequent reduction in the equation of state has been invoked in \cite{Carr:2019kxo,Carr:2023tpt} as a means to explain a peak in PBH production at this mass scale. However, \cite{Musco:2023dak} argue the impact of neutrino free streaming could negate the pressure reduction meaning that PBH production would not be enhanced. In either case, the reduction in pressure is much less significant than during the QCD transition, and even the QCD transition only leads to a reduction in the required power spectrum amplitude to generate a given PBH fraction by of order 10\%, compared to the value in a purely radiation dominated universe \cite{Byrnes:2018clq,Franciolini:2022tfm,Escriva:2022bwe,Musco:2023dak}.

The plan of our paper is as follows: In \cref{sec:G} we introduce some formalism and show the constraints on the PBH and $\mu$ distortions for Gaussian fluctuations. In \cref{sec:beyond-G} and \cref{sec:non-pert-nG} we extend the results to non-Gaussian fluctuations, starting with perturbative values of (local) $\fnl$ and then going to non-perturbative non-Gaussianity, including extreme forms of non-Gaussianity consisting of Gaussian perturbations raised to a large integer power. We discuss possible issues with invoking large non-Gaussianity as a means to evade the distortion constraints near the start of \cref{sec:non-pert-nG}. We conclude in \cref{sec:conclusion} and derive the relation between the PBH collapse fraction and the power spectrum amplitude for extreme forms of non-Gaussianity in \cref{sec:app-extreme}.

\section{Gaussian fluctuations: formalism and bounds on PBH abundance} \label{sec:G}

In this section, we first present the constraints on the PBH abundance in the SMBH range as well as the best possible PBH constraint which consists of 1 PBH in our cosmological horizon. We present Press-Schechter theory as a means to convert these constraints onto an amplitude of the primordial power spectrum. We also discuss the modelling uncertainties in our results.

\subsection{Observational PBH constraints}

There are a huge number of constraints on the fraction of dark matter which can consist of PBHs, parameterised by the dimensionless ratio $\fpbh$. For the supermassive black hole mass range, key constraints include X-ray and CMB distortion constraints from BH accretion\footnote{not to be confused with the distortion constraints on PBH formation studied in this paper.} and dynamical halo friction, see \cite{Carr:2020gox} for a review and references therein. Constraints are mass dependent and also depend on the PBH mass function but for a monochromatic (or narrow) mass spectrum are typically around $\fpbh\lesssim 10^{-4}$ \cite{Carr:2020xqk,Carr:2020erq,Deng:2021edw} (for model dependent constraints via magnetic fields see \cite{Papanikolaou:2023nkx,Papanikolaou:2023cku}). The constraints do not significantly change unless considering a very broad mass spectrum which would correspond to a very broad power spectrum and is not of great interest when attempting to evade distortion constraints. 

The abundance in PBHs today compared to the DM density ($f_{\rm PBH}$) is related to the fraction of the universe in PBHs at formation ($\beta$) via the simple relation (for a monochromatic mass spectrum) 
\begin{equation}
     f_{\rm PBH}\equiv \left. \frac{\rho_{\rm PBH}}{\rho_{\rm DM}}\right|_{\rm today}\simeq 2.4\beta \left(\frac{M_{\rm eq}}{M}\right)^{1/2}, \label{f-beta}
\end{equation}
where $M_{\rm eq}\simeq 2.7\times10^{17}M_\odot$ is the horizon mass at radiation-matter equality and the numerical factor arises because $f_{\rm PBH}$ is measured relative to dark matter and the ratio of DM at equality is $\rho_{\rm tot}/\rho_{\rm DM}=2(1+\Omega_{\rm baryon}/\Omega_{\rm DM})\simeq 2.4$ \cite{Planck:2018vyg}. This relation assumes the PBH mass will be constant after formation, whilst in reality accretion is expected to be significant for SMBHs, but it is very hard to estimate since it is a non-linear process \cite{Volonteri:2021sfo}.

Assuming that SMBHs do have a primordial origin, the minimum value of $\beta$ required to seed all SMBHs has not been quantified to the best of our knowledge, but the value is clearly going to be significantly smaller than the corresponding value of $f_{\rm SMBH}$ today related to $\beta$ using \cref{f-beta}, given that 1) quasars demonstrate that at least some SMBH accrete significantly and 2) SMBHs come in a large range of masses meaning that if they originated from a narrow PBH mass function then the heaviest SMBHs today gained most of their mass via accretion. Given these uncertainties, we often show constraints on $\fpbh=10^{-5}$ (by mapping back to a primordial value of the collapse fraction $\beta$ without including accretion) which is chosen as a value consistent with observational upper bounds on the PBH abundance yet still greater than the value which would be required to seed all SMBHs. 

Constraints on the $\mu$-distortion are often reported as a function of (inverse comoving) scale $k$ whilst those on PBHs as a function of mass, which we equate to the horizon mass at horizon entry ($M_H$) of $k$ using\footnote{We note this is quite crude since in reality, a spread of PBH masses will form and that there is not even an agreement on the relation between the peak PBH mass value and the horizon mass \cite{Niemeyer:1997mt,Yokoyama:1998xd,Musco:2008hv}, with estimates ranging by about a factor of 5 in either direction, see e.g.~\cite{Gow:2020bzo}. This uncertainty is important but outside the scope of this paper.}
\begin{equation}
    M_{\rm PBH}=M_H\simeq 17\left(\frac{g}{10.75}\right)^{-1/6}\left(\frac{k}{10^6\, {\rm Mpc}^{-1}}\right)^{-2} M_\odot, \label{M-k-relation}
\end{equation}
where $g$ is the number of relativistic degrees of freedom which we approximate as equal to $10.75$ for the rest of this paper since we are interested in PBH formation at temperatures well below the QCD scale.

Using Press-Schechter theory, the collapse fraction of the universe at PBH formation is given by
\begin{equation}
    \beta\equiv\frac{\rho_{\rm PBH}}{\rho_{\rm tot}}{\Big |}_{\rm formation}=\int_{\delta_c}^{\infty}P(\delta)d\delta \simeq \int_{{\cal R}_c}^{\infty}P(\calR)d\calR, \label{eq:PS}
\end{equation}
where $P$ is the probability density function (not to be confused with the power spectrum), $\delta$ is the density contrast and $\calR$ is the curvature perturbation. The values $\delta_c$ and ${\cal R}_c$ are PBH collapse thresholds that can be estimated from simulations, see \cite{Escriva:2021aeh} for a review and references therein. Going to the final expression in \cref{eq:PS} is expected to be a reasonable approximation in the case of a narrowly peaked power spectrum, such as we study here \cite{Young:2020xmk,Young:2022phe} (but see \cite{Germani:2023ojx}).

To be concrete, we assume a Dirac delta-function power spectrum
\begin{equation}
    {\cal P}_{\cal R}(k)=A k_* \delta(k-k_*),
    \label{eq:pr_dirac}
\end{equation}
with variance normalised to satisfy
\begin{equation}
    A=\sigma^2=\int_0^\infty \frac{dk}{k} {\cal P}_{\cal R}(k).
    \label{eq:pr_norm}
\end{equation}

We note that a delta function spike is of course unphysical, but it serves a useful purpose as the narrowest spike possible in principle, it simplifies analytical calculations, and it is the power spectrum shape which can generate the heaviest possible PBHs while remaining consistent with $\mu$ distortion constraints. Ref.~\cite{Gow:2020bzo} showed that both the required power spectrum amplitude and PBH mass function do not significantly vary when considering a fairly narrow lognormal mass function. In general, broad power spectrum peaks lead to broader constraints from $\mu$-distortions and hence restrict the formation of heavier PBHs more tightly, even though the amplitude required to form PBHs is slightly reduced.

Assuming Gaussian perturbations the pdf can be easily integrated to find
\begin{equation}
    \beta\simeq \frac12 \erfc\left(\frac{\calR_c}{\sqrt{2A }}\right) \label{beta-G}
\end{equation}
where $\erfc$ is the complementary error function. 
There have been many numerical and analytic calculations of the values of the collapse thresholds (for a discussion of how the collapse threshold depends on the density profile and how this is related to the power spectrum shape see e.g.~\cite{Harada:2013epa,Musco:2018rwt,Germani:2018jgr,Kalaja:2019uju}) 
and for definiteness we take $\calR_c=0.67$ in agreement with \cite{Nakama:2017qac}. More accurate techniques exist for calculating the PBH abundance, but we have checked that for this choice of $\calR_c$ the power spectrum amplitude required to generate a given value of $\fpbh$ (or equivalently $\beta$) matches the values derived in \cite{Gow:2020bzo}  to good accuracy but with far less computational cost.  

However, even if the curvature perturbations are precisely Gaussian the non-linear relation between the curvature perturbation and $\delta$ leads to a minimum possible level of non-Gaussianity in the density contrast (and compaction function, which is often used to study PBH formation). The impact of this non-linear relation is substantial and it results in the constraint on the power spectrum  being a factor of $2.0$ larger (weaker) than it would be if the linear relation was used (independently of the value of $\beta$) \cite{DeLuca:2019qsy,Young:2019yug,Gow:2020bzo}, and we take this factor into account in the constraints we show relating $\fpbh$ to the power spectrum amplitude by modifying \cref{beta-G} to
\begin{equation}
    \beta\simeq \frac12 \erfc\left(\frac{\calR_c}{\sqrt{A}}\right) \label{beta-GNL}
\end{equation} 
We assume this multiplicative factor applies in the same way when the curvature perturbations are intrinsically non-Gaussian, as done by \cite{Unal:2020mts}.

Assuming Gaussian perturbations, the constraint on the power spectrum amplitude is only logarithmically sensitive to $\beta$, and hence the amplitude required to form only 1 PBH inside the current cosmological horizon is not much smaller than the current constraint from the non-detection of PBHs. 
The constraint to have zero PBHs of scale $k$ today measured in terms of the $\beta$ parameter is \cite{Cole:2017gle} (see also \cite{Carr:1997cn,Nakama:2017xvq} where this is called the incredulity limit)
\begin{equation}    \beta<\betaz(k)\simeq1.2\times10^{-11}(k \, {\rm Mpc})^{-3}, \label{1-pbh}
\end{equation}    
where $\betaz$ denotes the PBH fraction at formation corresponding to a single PBH in today's cosmological horizon.

To give an order of magnitude estimate, forming 1 PBH with mass $10^6 M_\odot$ corresponds to $\betaz\simeq10^{-22}$ (and $\fpbh\simeq10^{-16}$) which is over 10 orders of magnitude tighter than the corresponding observational constraint on $\beta$, but the corresponding constraint on the power spectrum amplitude is only tighter by a factor of 2.3. Note that the 1-PBH limit corresponds to a constant number density and hence corresponds to  value of $\fpbh$ that depends on $M_{\rm PBH}$.

\subsection{Gaussian power spectrum constraints}

\begin{figure}[hbt!]
        \centering
        \includegraphics[width=\textwidth]{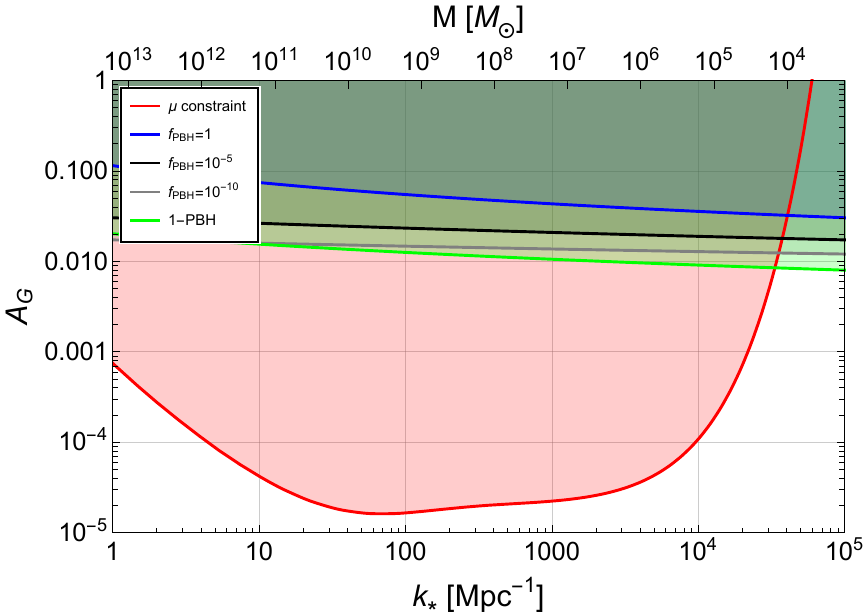}
        \caption{Constraints on the power spectrum amplitude assuming Gaussian curvature perturbations and a Dirac delta power spectrum. The upper x-axis shows the horizon mass in solar mass units (which is approximately the PBH mass) corresponding to each $k_*$ value.}
        \label{fig:G-const}
\end{figure}

Assuming the curvature perturbations are Gaussian but including the impact of the non-linear transformation between $\cal R$ and the density perturbation $\delta\rho$ \cite{DeLuca:2019qsy,Young:2019yug} we plot the PBH constraints for several values of $\fpbh$ and the single PBH \cref{1-pbh} limit, assuming a Dirac-delta function peak, in \cref{fig:G-const}. 

For the $\mu$ constraint, we use the accurate numerical techniques described in e.g.~\cite{Cyr:2023pgw,Tagliazucchi:2023dai} and summarized in our companion paper \cite{paper1} to calculate the power spectrum amplitude which saturates the COBE-FIRAS $\mu$-type distortion constraint. We use the updated value of $\mu<4.7\times10^{-5}$ of Bianchini and Fabbian \cite{Bianchini:2022dqh}, which is a factor 2 tighter than the constraint originally reported by the COBE collaboration \cite{Fixsen:1996nj} and slightly tighter than the TRIS result \cite{Gervasi:2008eb}.

The maximum PBH mass which can be generated by a peaked power spectrum and without overproducing spectral distortions is determined by the large $k$ value where the PBH and $\mu$-distortion constraints meet for any given power spectrum peak, and the corresponding horizon mass \cref{M-k-relation} gives an approximate estimate of the PBH mass which would form on this scale. One can see from \cref{fig:G-const} that there is no significant change in the value of $k$ (or $M_{\rm PBH}$) even when going from $\fpbh=1$ to generating just a single PBH in our entire observable universe. This is partly because $\beta$ is exponentially sensitive to the power spectrum amplitude, and partly because the $\mu$ constraint varies sharply with $k$ in the tail of the constraint. For the latter reason, the intersection is also not very sensitive to changes in the $\mu$ constraint. Broader power spectra lead to broader $\mu$ constraints and hence a lower possible maximum mass, see e.g.~\cite{Gow:2020bzo,Cyr:2023pgw}. 

\section{Beyond Gaussian fluctuations}\label{sec:beyond-G}

In this section, we show how to extend the calculation of the constraints to non-Gaussian curvature perturbations, but we first start by highlighting some of the challenges which non-Gaussian models face when being used as a means to generate PBHs.

One issue with PBHs formed from non-Gaussian perturbations is that they will form in clusters, and there are observational constraints on such clustering which may rule out evading $\mu$-distortion constraints for supermassive PBH generation via large non-Gaussianity, although the constraint will depend on the shape and amplitude of non-Gaussianity \cite{Shinohara:2021psq,DeLuca:2022uvz,Shinohara:2023wjd}.
There are also tight limits on any non-Gaussian correlation between PBH-forming scales and CMB scales via the constraint on photon-DM isocurvature perturbations \cite{Tada:2015noa,Young:2015kda,vanLaak:2023ppj}. These constraints are very tight when most of the DM is in PBHs but weaken significantly when $f_{\rm PBH}\ll1$, and such small values are  expected for the supermassive mass range. 
The scale corresponding to SMBH formation is only around 3 orders of magnitude smaller than CMB scales, so the standard assumption that these scales are uncoupled (even if the PBH scales are much more non-Gaussian) is not obviously true. Nonetheless, we here assume the usual CMB scales are sufficiently decorrelated from the large non-Gaussianity present on the PBH forming scales that this is not a problem, in which case the non-Gaussianity constraint from anisotropic $\mu$-distortions via $\mu-T$ correlations would also not apply \cite{Pajer:2012vz,Ganc:2012ae,Chluba:2016aln,Ravenni:2017lgw,Ozsoy:2021qrg,Zegeye:2021yml,Bianchini:2022dqh,Rotti:2022lvy,CMB-S4:2023zem}. In effect, we are assuming that $\fnl$ is scale-invariant over the small range of scales relevant for the large amplitude peak responsible for forming PBHs, but that its value varies strongly between the peak and the usual larger CMB scales. The clustering and isocurvature constraints will be irrelevant in the limit of very few PBHs (because this makes clustering impossible) but may be important when a primordial seed for all or most SMBHs is being considered.

When considering very non-Gaussian tails to the pdf we should consider the impact of type II perturbations (those with $\delta$ significantly larger than $\delta_c$ whose formation has recently been simulated \cite{Uehara:2024yyp}), which have been claimed to form separate universes rather than PBHs though \cite{Kopp:2010sh} argues against this.
For (approximately) Gaussian perturbations the type II perturbations are exponentially suppressed compared to the usual type I perturbations which form PBHs and are hence irrelevant, but if the tails are sufficiently flat then they may become relevant \cite{Gow:2022jfb}.

\subsection{Local non-Gaussianity: Perturbative and non-perturbative limits}

Primordial non-Gaussianity of the local type, which arises when deviations from Gaussianity are local in real space, is usually characterised by a Taylor expansion of the comoving curvature perturbation about its mean 
\begin{equation}    
{\cal R}(\vec{x})=
    {\cal R}_{\rm G}(\vec{x})
    + \fnl 
    \left( {\cal R}_{\rm G}(\vec{x})^2
    - \langle{\cal R}_{\rm G}(\vec{x})\rangle^2
    \right) + \gnl 
     {\cal R}_{\rm G}(\vec{x})^3
     + \cdots,
    \label{eq:Rg}
\end{equation}
where $\fnl\equiv 3f_{\rm NL}/5 $ and $\gnl\equiv 9g_{\rm NL}/25$ are non-linearity parameters quantifying the magnitude of non-Gaussianity. We denote the variance of curvature as $A \equiv \langle {\cal R}^2 \rangle$ and that of the underlying Gaussian field as $A_{\rm G} \equiv \langle {\cal R}_{\rm G}^2 \rangle$.

Perturbative non-Gaussianity means that the linear term dominates, i.e.~$|\fnl| A_{\rm G}^{1/2} \ll 1$, and $|\gnl |A_{\rm G} \ll 1$. Since in this limit $A$ and $A_{\rm G}$ are almost equal, these conditions are equivalent to $|\fnl| A^{1/2} \ll 1$, and $|\gnl |A \ll 1$.  Non-perturbative non-Gaussianity implies the contrary, with either the $\fnl$, $\gnl$, or even higher-order terms dominating to the extent that the power spectrum is completely dominated by this non-linear term.

 In most of this paper, we assume that $\gnl$ is negligible and focus on the non-Gaussianity arising from $\fnl$. We focus on $\fnl>0$ because this corresponds to positive skewness which boosts PBH production, but we note that unlike the extremely asymmetric response of PBH constraints to the sign of the non-Gaussianity \cite{PinaAvelino:2005rm,Lyth:2012yp,Byrnes:2012yx}, the $\mu$ constraints depend primarily on the magnitude of $\fnl$ \cite{paper1}. We discuss $\gnl$ in \cref{sec:gNL} and \cref{sec:app-extreme} and extend the expansion of \cref{eq:Rg} to higher orders in \cref{sec:higher-nG}.

\subsection{Perturbative non-Gaussianity constraints}\label{sec:pert-nG}

Here we show constraints both in the truly perturbative limit of $\fnl \sqrt{A}\ll1$ and the borderline case of perturbativity ($\fnl \sqrt{A}=1$) where the perturbative treatment is no longer really valid, but we include this to give an estimate of the maximum possible impact of `perturbative' non-Gaussianity. 

Using the fact that $\langle {\cal R}_G^4\rangle=3\langle {\cal R}_G^2\rangle^2=3A_G^2$, \cref{eq:Rg} shows that the total variance is given by
\begin{equation}
    A\equiv\langle {\cal R}^2\rangle = A_G+2\fnl^2 A_G^2. \label{eq:variance}
\end{equation}

The spectral distortion for an arbitrary value of $\fnl$ for a spectrum centred at the scale $k_*$ is given by 
\begin{equation}
    \mu=\mu_G+\mu_{\chi^2}=A_G \Fone(k_*)+\fnl^2 A_G^2 \Ftwo(k_*)~.
\end{equation}
More details are given in paper I, including the definitions of $\Fone$ and $\Ftwo$ and useful fitting functions \cite{paper1}. We can then compute the variance $A$ which saturates the observational $\mu$ constraint for each chosen value of $\fnl^2 A_{\rm G}$ or $\fnl^2 A$.

The technique for determining the PBH constraints with arbitrary values of $\fnl$ is detailed in paper \cite{Byrnes:2012yx} and explained for certain limiting cases in \cref{sec:app-extreme}. Following those techniques we first work in terms of the variance $A_{\rm G}$ of the Gaussian perturbations, since the relevant formulas for $\fpbh$ are defined in terms of that variable \cite{Byrnes:2012yx}\footnote{We note that modern techniques to determine the PBH abundance subject to non-Gaussianity have found the impact of finite $\fnl$ is significantly reduced compared to the calculation of \cite{Byrnes:2012yx}, see e.g.~\cite{Young:2022phe,Ferrante:2022mui}, but this would not change our main conclusions.}. We then convert this result into a total variance $A$ using \cref{eq:variance} and the fact we want to set $\fnl$ by the condition 
\begin{equation}    
\fnl^2\,A=\kappa \, ,
\end{equation}
for a given value of $\kappa$, which quantifies the relative non-Gaussian contribution to the total variance. In the next figure, we display our results for $\kappa=0.01$ and $\kappa=1$\footnote{Note that in paper 1 \cite{paper1} we instead plot results for fixed values of $\fnl^2\,A_G$.}. Plugging this value of $\fnl$ into \cref{eq:variance} we find the variance to be related to the variance of the Gaussian perturbations by 
\begin{equation}
    A=\frac{A_G}{2}\left(1+\sqrt{1+8\kappa}\right).
\end{equation}
Hence for $\fnl^2A=0.01$ the variance only varies from the amplitude of the Gaussian power spectrum at the percent level, but for the borderline perturbative case of $\fnl^2 A=1$ we have $A=2A_G$, corresponding to equal contributions from the Gaussian and $\chi^2$ parts of the curvature perturbation.

\begin{figure}
    \centering
    \includegraphics[scale=1]{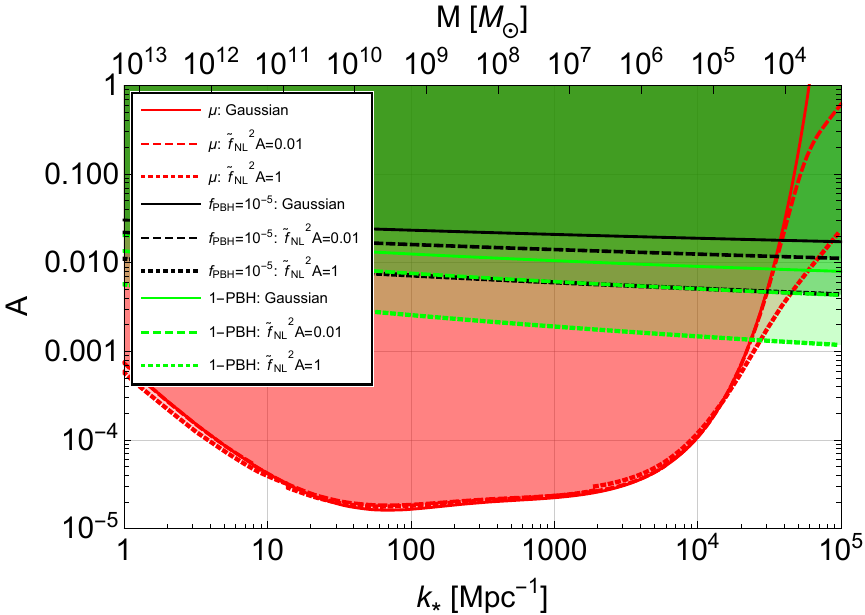}
    \caption{The $\mu$ and PBH constraints for Gaussian perturbations (solid), perturbative non-Gaussianity with $\fnl^2 A=0.01$ (dashed) and the limit with the maximum possible perturbative non-Gaussianity (dotted). Coincidentally, the PBH constraint with $\fpbh=10^{-5}$ and $\fnl^2 A=1$ (dashed black) is almost identical to the constraint with 1 PBH inside today's horizon and $\fnl^2 A=0.01$ (dotted green).  When including non-Gaussianity we note that only the Gaussian part of the power spectrum has a Dirac delta-function power spectrum. 
}
    \label{fig:P-perturbativeVarDash}
\end{figure}

\Cref{fig:P-perturbativeVarDash} shows how the distortion and PBH constraints change for both of these limits, where we have varied $\fnl$ as a function of $k_*$ (the position of the peak) such that it always takes the value consistent with the value of $\kappa$ shown. The first thing to notice from the figure is that the $\mu$ constraints remain much stronger than the PBH constraints over most of the scales where distortion constraints exist. To determine the maximum mass PBH which can be generated consistently with the $\mu$-distortion constraint one should determine the intersection of the appropriate PBH and $\mu$-constraint lines (solid with solid for Gaussian, dotted with dotted for $\fnl^2A=1$, etc). For Gaussian perturbations, the maximum mass is $\sim10^4M_\odot$ which is very similar to the maximum mass possible with $\fnl^2A=1$, but one would have overestimated the mass by a factor of 2 if one compared the correct PBH constraint to the Gaussian $\mu$ constraint. 

\section{Non-perturbative non-Gaussianity}\label{sec:non-pert-nG}

Having seen that even the maximum possible ``perturbative'' non-Gaussianity is insufficient to substantially weaken the PBH constraints, in this section we study the limit of completely non-perturbative non-Gaussianity. We first focus on the more commonly considered case of chi-squared non-Gaussianity for which we make a detailed calculation. We then sketch the even more extreme case of Gaussian cubed statistics and beyond. Before making these calculations, we outline some general issues with large non-Gaussianity and also comment that in practice the non-Gaussian perturbations act like a linear ``background'' perturbation in any finite volume, meaning that the Gaussian (linear) term will not be completely absent in practise \cite{Boubekeur:2005fj,Nelson:2012sb,Nurmi:2013xv,Young:2014oea}. But it's still useful to consider this completely non-Gaussian limit as an extreme scenario.

\subsection{Distortions in the limit of local $\chi^2$ non-Gaussianity}\label{sec:chi-sq}

In this section we focus on a pure chi-squared non-Gaussianity, meaning that we can absorb $\fnl$ into the power spectrum amplitude and write
\begin{equation}
    {\cal R}(\vec x)={\cal R}_{\rm G}^2(\vec x)-\langle{\cal R}_{\rm G}^2\rangle.
\end{equation}
Provided that the Gaussian power spectrum has a delta-function peak, one can determine the analytic shape and amplitude of the chi-squared power spectrum to be (as explained in paper 1 \cite{paper1}) 
\begin{equation} {\cal P}_{{\cal R}_G^2}(k)\equiv \frac{k^3}{2\pi^2}P_{{\cal R}_G^2}(k)= A_G^2 \frac{k^2}{k_*^2} H(2k_*-k)=A_{G^2}\frac{1}{2} \frac{k^2}{k_*^2} H(2k_*-k)~, \label{PS-chi-sq}
\end{equation}
where $H$ is the Heaviside function. The variance of this power spectrum is given by
\begin{equation}
   A_{G^2}= \int_0^\infty {\cal P}_{{\cal R}_G^2} \,\, d\ln k =  2 A_G^2,
\end{equation}
in agreement with the large $\fnl$ limit of \cref{eq:variance}.

\begin{figure}
    \centering
    \includegraphics[scale=1]{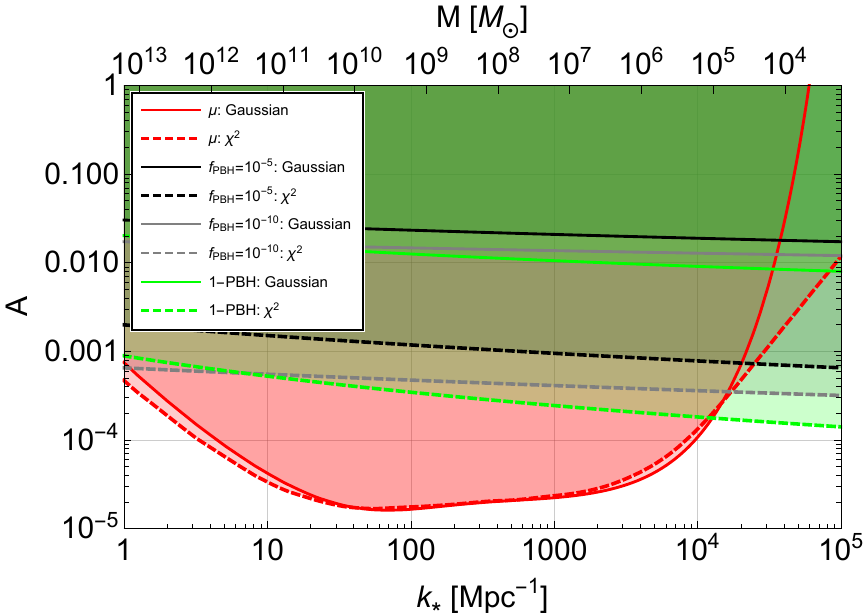}
    \caption{The $\mu$-distortion and PBH constraints on the variance for Gaussian (solid) and $\chi$-squared (dashed) statistics.}
    \label{fig:chi-sq}
\end{figure}

The $\mu$-distortion can then be computed using techniques described in \cref{sec:pert-nG} and paper I. The constraints from this power spectrum variance are shown in \cref{fig:chi-sq}. Notice that the $\mu$ constraint is normally similar to the Gaussian case, with the difference being primarily towards the tails where the constraints tighten, due to the shape of the power spectrum varying between the Gaussian and $\chi^2$ cases. 
The $\chi^2$ peak of \cref{PS-chi-sq} is asymmetric and broader to the left of the peak, which explains why the difference in the constraints becomes significant for $k\gtrsim 10^4\,{\rm Mpc}^{-1}$ where it is mostly the left-hand tail of the power spectrum peak which generates a $\mu$-distortion. 
The PBH constraint tightens as expected (for all PBH masses), demonstrating that PBH formation is for a large mass range more sensitive to non-Gaussianity than the $\mu$ constraints. However, the maximum mass PBH possible (i.e.~the intersection of the $\mu$ and PBH constraints) does not change very significantly for $\fpbh=10^{-5}$ (it increases by a factor of a few), because of the broader $\mu$-distortion constraint in the $\chi^2$ case. However, for smaller values of $\fpbh$ the difference becomes more important and for the extreme 1-PBH limit the largest possible mass with $\chi^2$ statistics is almost an order of magnitude larger than for Gaussian perturbations.

\subsection{General values of $\fnl$}

In this section we collate the perturbative and non-perturbative results for any value of $\fnl>0$, summarising some of the key information shown in \cref{fig:P-perturbativeVarDash,fig:chi-sq} in another way. 

\begin{figure}
    \centering
    \includegraphics[scale=1]{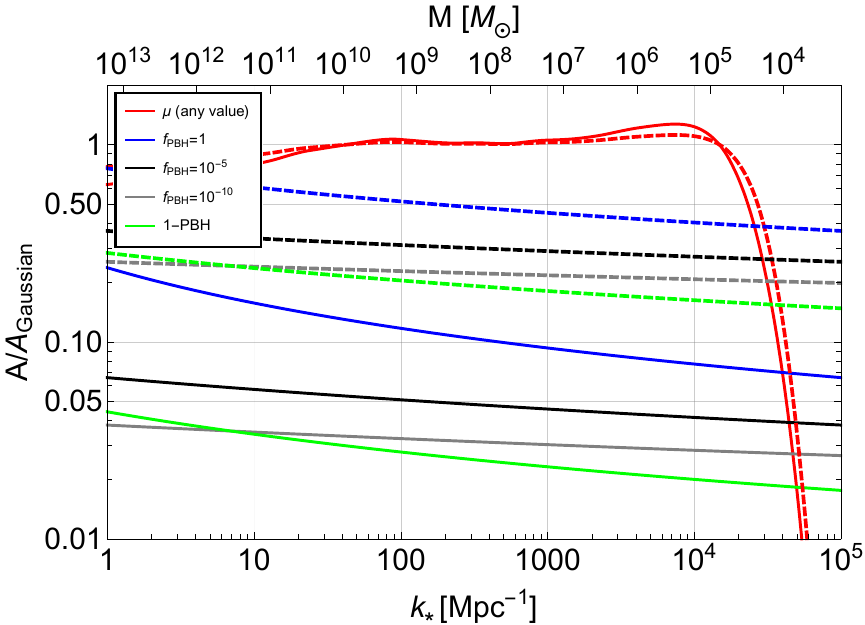}
    \caption{
    The ratio of the variance, comparing the constraints of the variance of either a pure $\chi^2$ perturbation to Gaussian (solid lines) or `perturbative' non-Gaussianity with $\fnl^2A=1$ compared to the Gaussian variance (dashed lines, all following the same colours). Note that the ratio for the $\mu$ constraint is independent of the value of $\mu$. In contrast, the PBH constraints do depend on the assumed value of $\fpbh$ but are only mildly dependent on $k_*$.  \label{fig:extremeRatio}
    }
\end{figure}

In \cref{fig:extremeRatio}, we plot the ratio of the variances leading to the same value of $\mu$ or $f_{\rm PBH}$, first, for the chi-squared case compared to the Gaussian case (solid lines), and second, for the $\fnl^2A=1$ case compared to the Gaussian case (dashed lines). We recall that the condition $\fnl^2A=1$ roughly sets the boundary between the perturbative and non-perturbative non-Gaussianity regime. For $\mu$-distortions,
the ratio is independent of the observed value of $\mu$ because $\mu$ depends linearly on the variance, and the ratio is close to unity except towards the tails in $k$ where the $\mu$ constraints weaken. In contrast, the ratio of the PBH constraints is always significantly less than one and is (mildly) dependent on the assumed constraint on the PBH fraction. The fact that the lines for a pure $\chi^2$ non-Gaussianity are always significantly below the equivalent lines for the perturbative limit non-Gaussianity demonstrates that it is only in the extremely non-Gaussian regime that the PBH constraints change by more than a factor of a few.

\subsection{Distortions from large Gaussian cubed non-Gaussianity}\label{sec:gNL}

Since even the non-perturbative limit of  $\chi^2$ non-Gaussianity is insufficiently non-Gaussian to make supermassive PBH formation compatible with the $\mu$-constraint, there is a substantial literature on more extreme forms of non-Gaussianity, e.g.~\cite{Nakama:2016kfq,Nakama:2017xvq,Unal:2020mts,Hooper:2023nnl}. The obvious next step is Gaussian cubed statistics, 
\begin{equation}
    {\cal R}={\cal R}_G^3, 
\end{equation}
corresponding to infinitely large $\gnl$.

An analytic estimate of the variance required for PBH formation follows from the techniques in \cite{Byrnes:2012yx}. Using the fact that $\langle {\cal R}_G^6\rangle=5\times3\langle {\cal R}_G^2\rangle^3=15A_G^3$, we see that in the limit of $\gnl$ completely dominating,
and as derived in \cref{sec:app-extreme}, the variance is related to the collapse fraction $\beta_{{\rm G}^{3}}$ by
\begin{equation}
    A_{G^3}=15 A_G^3=\frac{15}{8}\frac{\calR_c^2}{\inverfc^3(2\beta_{{\rm G}^{3}})}\,.
\end{equation}

\begin{figure}
    \centering
    \includegraphics[scale=1]{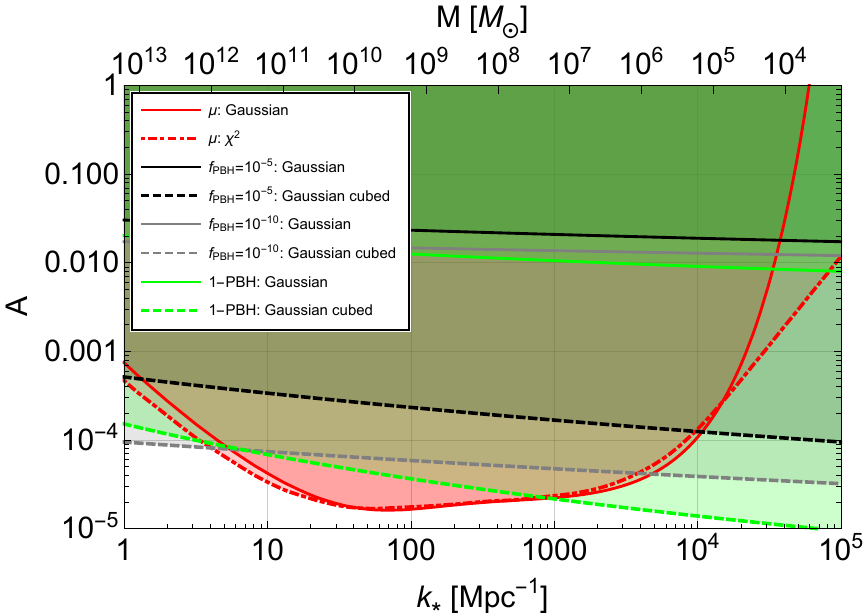}
    \caption{The $\fpbh$ constraints computed with non-perturbative Gaussian cubed perturbations (dashed lines), with the corresponding Gaussian constraints (solid) shown for comparison. The Gaussian and $\chi^2$ (dot-dashed) $\mu$-constraints are also plotted for comparison purposes.} \label{fig:pure-gnl}
\end{figure}

We show the corresponding constraints in \cref{fig:pure-gnl}. Unfortunately, we are unable to determine the variance corresponding to the $\mu$ constraint with Gaussian cubed non-Gaussianity, but expect it to be comparable in amplitude to the Gaussian and $\chi^2$ constraints, which are comparable to each other over a large range of $k$.  This shows that even cubic non-Gaussianity is at best borderline sufficiently non-Gaussian to allow the generation of even a single PBH inside the observable universe. The largest possible PBH looks likely to be only a factor of a few larger than for the Gaussian and $\chi^2$ cases for $\fpbh=10^{-5}$, but could potentially be two orders of magnitude larger for the extreme (and rather academic) limit of 1 PBH inside our universe. However, what is clear is that even in this case $f_{PBH}$ must be too small to explain the origin of all SMBHs. 

\subsection{Higher-order non-Gaussianity}\label{sec:higher-nG}

Since even the infinite limits of $\fnl$ and $\gnl$ are insufficiently non-Gaussianity to allow PBH formation with very large masses, we here consider a general curvature perturbation of the form 
\begin{equation}
    {\cal R}={\cal R}_G^n-\langle {\cal R}_G^n\rangle . \label{eq:R-n}
\end{equation}
but we caution that such extreme forms are not well motivated and that even if a model can be found\footnote{for a concrete model with $n=2$ see \cite{Gow:2023zzp}.}, having all terms with smaller $n$ set to zero is expected to require substantial fine-tuning \cite{Boubekeur:2005fj,Nelson:2012sb,Nurmi:2013xv,Young:2014oea}.
For positive integer $n$ we can estimate the values of the power spectrum, bispectrum and trispectrum in terms of the variance of the Gaussian perturbation ($A_G$) to be
\begin{equation}
    {\cal P}_{\cal R}\sim A_G^{n}, \qquad {\cal B}_{\cal R}\sim A_G^{3n/2} \; {\rm for\; even}\; n, \qquad {\cal T}_{\cal R}\sim A_G^{2n} \; {\rm for\;any}\; n,
\end{equation}
whilst the bispectrum would be zero for odd $n$. In any case, the largest contribution to these spectra is always the power spectrum whilst the non-Gaussian contributions are suppressed by a factor of $\sigma_G=A_G^{1/2}$ to a positive power, which must always be much less than unity in order to avoid over producing PBHs.
Hence, the tightest constraints on the amplitude of the variance coming from the $\mu$ distortion in the $k$ range of primary interest, which corresponds to $10\,{\rm Mpc}^{-1}\lesssim k\,  \lesssim\, 10^4{\rm Mpc}^{-1}$ is expected to be similar for any value of $n$. We have explicitly shown this to be true for the cases of $n=1$ and $n=2$. The tails to small and large $k$ differ because the shape of the power spectrum peak is a function of $n$, but unfortunately for $n>2$ the convolution integrals which should be computed in order to determine the power spectrum shape are too complicated to solve.

\begin{figure}
    \centering
    \includegraphics[scale=1]{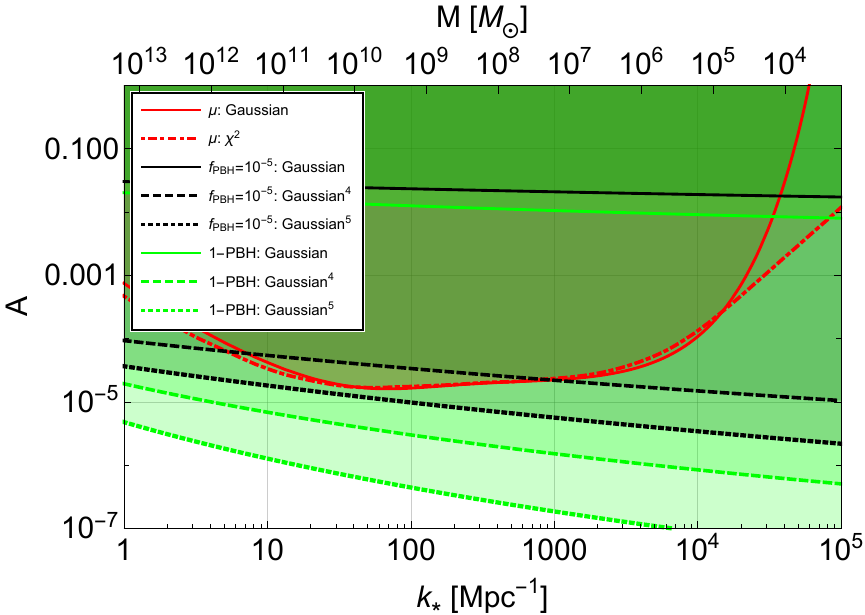}
    \caption{The $\fpbh$ constraints computed with non-perturbative Gaussian to the power 4 (dashed) or power 5 (dotted) perturbations, with the corresponding Gaussian (solid) values shown for comparison. The Gaussian and $\chi^2$ (dot-dashed) $\mu$ constraints are also plotted for comparison purposes.}
    \label{fig:G45}
\end{figure}

The PBH constraints for the extreme cases of $n=4$ and $n=5$ are shown in \cref{fig:G45}. This is based on calculations presented in \cref{sec:app-extreme}. 
For $n=4$ the maximum PBH mass increases substantially compared to the more Gaussian cases, with $M\gtrsim10^7M_\odot$ becoming realisable assuming $\fpbh=10^{-5}$. For even smaller values of $\fpbh$, it becomes possible to generate some PBHs of any mass. For $n=5$ the $\mu$ constraint becomes weaker than the PBH constraint even for $\fpbh>10^{-5}$.

\subsection{Literature comparison}

Here we compare our results to some other literature which studied the possibility of using large non-Gaussianity to evade the $\mu$ constraints. 
Our results for the PBH constraints are most similar to those of Unal et al \cite{Unal:2020mts}, who considered the infinite limits of $\fnl$ and $\gnl$ and also included the non-linear relation between the curvature perturbation and density perturbation. Our constraint curves are very similar, with the main difference arising because they use a collapse threshold of $0.5$, making their constraints on the power spectrum tighter than ours by a factor of $(0.5/0.67)^2\simeq0.56$. Note from \cref{sec:app-extreme} that the constraint on the variance is proportional to $\calR_c^2$ even in the case of highly non-Gaussian fluctuations. 

Nakama et al.~\cite{Nakama:2016kfq,Nakama:2017xvq} and Hooper et al.~\cite{Hooper:2023nnl} include a study of the following phenomenological parametrisation of the pdf 
\begin{equation}
 P(\calR)=\frac{1}{2\sqrt{2}\tilde{\sigma}\Gamma(1+1/p)}e^{-(|\calR|/(2\tilde{\sigma}))^p}, \label{P-p}
 \end{equation}
which has been normalised to satisfy $\int_{-\infty}^\infty P(\calR)d\calR=1$, where the variance of the curvature perturbation $\calR$ is given by
\begin{equation}
\sigma^2\equiv \int^{\infty}_{-\infty}\calR^2P(\calR)d\calR=\frac{2\Gamma(1+3/p)}{3\Gamma(1+1/p)}\tilde{\sigma}^2. 
\end{equation}
Note that for $p=2$ the perturbations are Gaussian and $\sigma=\tilde{\sigma}$, as expected. For $p<2$ the tail is flattened but the derivative of $P(\calR)$ is discontinuous at $\calR=0$, making this pdf unphysical. The case with
$p=1$ has the same tail to large $\calR>0$ as the $\chi^2$ distribution but is not equivalent (with \cref{P-p} being symmetric, unlike the $\chi^2$ distribution). 
Smaller values of $p$ correspond to flatter tails and hence an enhanced PBH abundance. 
The advantage of this parameterization of the curvature perturbation is that it is straightforward to calculate the PBH abundance using Press-Schechter theory, within the limitation that one needs to use approximate methods for determining the PBH abundance using the curvature perturbation directly. 
%

Taking results from \cref{sec:app-extreme}, from \cref{PNG} and \cref{y-formula} one can find a relation between the extreme non-Gaussian curvature perturbation of \cref{eq:R-n} and the pdf of \cref{P-p}. With odd $n$ the tail of the pdf goes like $2/n$ which hence corresponds to $p=2/n$ in the tail of \cref{P-p}. For even $n$ there is no such simple relation in general, but one can see from \cref{y-G2} that $n=2$ corresponds to having the same pdf tail as $p=1$. The pdfs are not equivalent except in the large $\calR$ tails, but this is the part of the pdf which is relevant for PBH formation.

Based on our result that a Gaussian cubed ($n=3$) perturbation is not sufficiently non-Gaussian to generate primordial SMBHs whilst evading the $\mu$ constraint, while Gaussian to the fifth power is sufficiently non-Gaussian suggests that one requires $p$ to be less than somewhere in the range $2/5-2/3$. This agrees with the result of Nakama et al.~\cite{Nakama:2017xvq}, who suggest the threshold is around $p=0.5$. Compared to them, on the one hand, we use a $\mu$ constraint a factor of 2 tighter, but on the other hand, we include the unavoidable non-Gaussianity caused by the non-linear relation between the curvature perturbation and density contrast, which weakens the PBH constraint on the variance by a factor of 2. Thus our results should indeed be similar. 

Hooper et al.~\cite{Hooper:2023nnl} suggest a similar constraint on $p$ and propose a curvaton model with a non-quadratic potential as a means to generate a sufficiently non-Gaussian perturbation including a kinetic coupling to the inflaton to generate the required peak in the power spectrum, but do not find a concrete model which works. 

The case of pure $\chi^2$ statistics \cref{sec:chi-sq} leads to the exponential tail corresponding to $p=1$. This behaviour in the tail of the pdf has been motivated by numerous authors studying stochastic inflationary effects, e.g.~generated during ultra-slow-roll inflation in some cases \cite{Pattison:2017mbe,Ezquiaga:2019ftu,Figueroa:2020jkf,Achucarro:2021pdh,Cai:2022erk,Gow:2022jfb,Pi:2022ysn}. Our results show that these exponential tails are insufficiently non-Gaussian to generate primordial SMBHs whilst evading the $\mu$ constraints. The observation that $\chi^2$ statistics reduce the required power spectrum variance by an order of magnitude (see \cref{fig:chi-sq}) is consistent with the stochastic inflation analysis of \cite{Tomberg:2023kli}. From \cref{fig:G-const} one can infer that one needs a reduction in the variance by about 3 orders of magnitude to generate PBHs of any mass and in this paper, we have argued that this inference will remain approximately true even when the distortion constraint is recalculated to include the impact of non-Gaussianity. Of course, the fact that the pdfs only agree in the tail means the required variance to generate a given abundance of PBHs will not be exactly the same if the pdf is not exactly the same everywhere, but we have checked that the difference between the pure $\chi^2$ and $p=1$ pdfs is of order $10\%$, which is insignificant compared to the orders of magnitude gap between the PBH and distortion constraints as shown in \cref{fig:chi-sq}.

\section{Conclusions}\label{sec:conclusion}

As is well known, the tight constraint on cosmic $\mu$-distortions rules out the formation of supermassive PBHs, assuming Gaussian perturbations. Given that the origin of SMBHs remains a mystery, and that they are even observed at high redshift, numerous efforts have been made to invoke sufficiently extreme forms of non-Gaussianity in order to evade the $\mu$-constraints and allow the formation of PBHs with any mass \cite{Nakama:2016kfq,Nakama:2017xvq,Unal:2020mts,Juan:2022mir,Hooper:2023nnl}. However, none of these papers have recalculated the $\mu$-distortion constraint which means their conclusions may not be correct. 

In our companion paper \cite{paper1} we have made the first full calculation of the $\mu$-distortion subject to local non-Gaussianity parameterised by $\fnl$, and in this paper we have made the first comparison between the corresponding PBH and $\mu$-constraints, showing how the respective constraints change when dropping the assumption of Gaussian perturbations. We note that many previous papers have considered the correlation of $T$ and $\mu$-distortion perturbations as a means to constrain non-Gaussianity on small scales, but they have not calculated the averaged background value of $\mu$ \cite{Pajer:2012vz,Ganc:2012ae,Khatri:2015tla,Chluba:2016aln,Ravenni:2017lgw,Ozsoy:2021qrg,Zegeye:2021yml,Bianchini:2022dqh,Rotti:2022lvy,CMB-S4:2023zem} and we also assume that the large non-Gaussianity does not correlate to CMB scales.

Because PBH formation takes place deep in the tail of the pdf, the formation rate is highly sensitive to non-Gaussianity and the required power spectrum amplitude can change significantly (by more than an order unity correction) even for perturbative levels of non-Gaussianity, whilst the $\mu$ constraints are primarily sensitive to the peak of the pdf and hence do not change significantly. However, towards the large $k$ tail of the range of scales which distortions can constrain the $\mu$ constraint tightens significantly, reducing the maximum mass with which any PBH can form. Hence -- in the interesting limit of highly non-Gaussian perturbations -- it is incorrect to neglect the change (the strengthening) of the $\mu$-distortion constraint. 

For a pure $\chi^2$ non-Gaussianity we are able to perform a full calculation of both the $\mu$ and PBH constraints and find that the maximum mass with which PBHs can be generated remains comparable to the case of Gaussian perturbations, no matter which value of $\fpbh$ is desired. Our tentative results for a Gaussian cubed perturbation show that such statistics may be close to sufficiently non-Gaussian to generate a tiny fraction of supermassive PBHs of any mass, but at a level insufficient to explain the origin of all SMBHs. 

Unfortunately, the techniques we have developed to determine the $\mu$ distortion constraint for $\chi$-squared non-Gaussianity cannot easily be extended to more extreme forms of non-Gaussianity so we cannot definitively determine which more extreme form of non-Gaussianity might allow primordial SMBH production of any mass. However, one can see from \cref{fig:chi-sq,fig:extremeRatio} that the tightest constraint on the variance of the perturbations due to the $\mu$ distortion, which applies over a large range of scales corresponding to the mass range $10^5M_\odot\lesssim M\lesssim 10^{11} M_\odot$, barely changes. We explain the reasons carefully in our companion paper and expect this result to remain true for more extreme forms of non-Gaussianity. Using results developed in the appendix, we extend the calculation of the PBH constraints to Gaussian to higher powers than squared and conclude from \cref{fig:G45} that a Gaussian perturbation raised to the fifth power would in principle be sufficiently non-Gaussian to allow the formation of a significant fraction of PBHs of any mass without conflicting with the distortion constraints.

We caution that even if a working model with large non-Gaussianity can be found which evades the $\mu$-distortion constraints, and if an actual inflationary model to generate can be constructed, the other challenges described near the start of \cref{sec:non-pert-nG} remain. In addition, there are potentially tighter but less well-understood and model-dependent constraints coming from dark matter substructure, including ultracompact minihaloes, on a more limited range of scales, see e.g.~\cite{Bringmann:2011ut,Shandera:2012ke, Gosenca:2017ybi,Nakama:2017qac,Karami:2018qrl,Delos:2018ueo,Ando:2022tpj} and a constraint from BBN \cite{Jeong:2014gna}. 

Throughout this paper, we have assumed PBHs form via the direct collapse of large amplitude density perturbations shortly after horizon entry. Alternative formation mechanisms exist (for examples generating supermassive PBHs see e.g.~\cite{Deng:2017uwc,Huang:2023mwy,Kasai:2024tgu}) and the calculation of the $\mu$ constraint would have to be redone for each scenario. However, PBH formation in all these cases takes place later than horizon entry (meaning on subhorizon scales relevant for CMB distortions and still on a scale comparable to the PBH which forms), so one should not - in general - expect these cases to easily evade the tight CMB spectral distortion constraints which apply during PBH formation.

{\bf Acknowledgements} The authors thank Jens Chluba, Aurora Ireland, Xav Pritchard, David Seery, Caner Unal and Sam Young for helpful conversations.  
We thank the Royal Society for funding an International Exchanges Scheme which allowed the three authors to travel between Brighton and Aachen. CB is supported by STFC grants ST/X001040/1 and ST/X000796/1. DS is supported by the German Academic Exchange Service (DAAD) grant.

\appendix 

\section{Calculating the PBH abundance with extreme non-Gaussianity}\label{sec:app-extreme}

We here summarise the calculation which determines the power spectrum amplitude as a function of $\beta$ (or $\fpbh$) for non-Gaussian perturbations. We follow \cite{Byrnes:2012yx} and refer the reader to that paper for the calculation with finite values of $\fnl$ or $\gnl$. We here focus on the simpler cases of the extreme non-Gaussian limit of \cref{sec:higher-nG}
\begin{equation}
    \calR\equiv h(\calRG)= \calRG^n-\langle\calRG^n\rangle,
\end{equation}
for positive integer $n$. To the best of our knowledge we extend this calculation to $n=4$ and $n=5$ for the first time. These correspond to the infinite limits of $h_{\rm NL}$ and $i_{\rm NL}$ respectively \cite{Young:2013oia}. We reiterate that large non-Gaussianity needs to be treated with caution, see \cref{sec:beyond-G}. 

The crucial idea is to transform from the non-Gaussian (NG) probability density function (PDF) to a Gaussian PDF using the following transformation 
\begin{equation}
    P_{\rm NG}d\calR = \sum_i {\bigg |}\frac{d h_i^{-1}(\calR)}{d\calR}{\bigg |} \, P_{\rm G}(h^{-1})d\calR , \label{PNG}
\end{equation}
where $P_{\rm G}$ has a Gaussian distribution. The sum is over all solutions of the equation $h(\calRG)=\calR$. For even values of $n$ there are always two identical solutions with $\calR>\calR_c$ and for odd $n$ there is always one solution.

We start by considering odd $n$, which is the simpler case since in this case $\langle\calRG^n\rangle=0$. Therefore, $h^{-1}(\calR)=\calR^{1/n}$ and changing variables to
\begin{equation}
    y\equiv \frac{h^{-1}(\calR)}{\sG}= \frac{\calR^{1/n}}{\sG}, \qquad y_c=\frac{\calR_c^{1/n}}{\sG}, \label{y-formula}
\end{equation}
one can use Press-Schechter theory to find
\begin{equation}
    \beta_{{\rm G}^{ n}} =\frac{1}{2} \erfc\left(\frac{y_c}{\sqrt{2}}\right).  
\end{equation}
We invert this to find
\begin{equation}
A_G=\sG^2=\frac{\calR_c^{2/n}}{2 \inverfc^2(2\beta_{{\rm G}^{n}})},    
\end{equation}
where $\inverfc$ in the inverse of the complementary error function and hence the variance (for odd $n$) satisfies
\begin{equation}
    (n-1)!!A_{\rm G}^{n}=\frac{(n-1)!!}{2^n}\frac{\calR_c^2}{\inverfc^n(2\beta_{{\rm G}^{n}})}\,.
\end{equation}

For even $n$ the subtraction of the $\langle\calRG^n\rangle$ term means we cannot derive a general expression, but we here derive the solution for $n=2$ and 4, starting with the case of pure $\chi^2$ non-Gaussianity, for which $\langle\calRG^2\rangle=\sG^2$. We here have $h^{-1}=\pm \sqrt{\calR+\sG^2}$ and hence using the variables 
\begin{equation}
    y= \frac{\pm \sqrt{\calR+\sG^2}}{\sG}, \qquad y_c=\frac{\pm \sqrt{\calR_c+\sG^2}}{\sG}, \label{y-G2}
\end{equation}
one finds
\begin{equation}
    \beta_{{\rm G}^2}=\erfc\left(\frac{|y_c|}{\sqrt{2}}\right), \qquad \Rightarrow \,\, |y_c|=\sqrt{2}\, \inverfc\left(\beta_{{\rm G}^2}\right),
\end{equation}
which we can rearrange to find the variance is 
\begin{equation}
   A_{G^2}= 2\sG^4=\frac{2\,\calR_c^2}{\left(2\,\inverfc^2\left(\beta_{{\rm G}^2}\right)-1\right)^2}.
\end{equation}

Finally, for $n=4$, and using $\langle\calRG^4\rangle=3\sG^4$, we have $h^{-1}=\pm\left(\calR+3 \sG^4\right)^{1/4}$, and changing to the same variables as before, 
\begin{equation}
    y= \frac{\pm \sqrt{\calR+3\sG^4}}{\sG}, \qquad y_c=\frac{\pm \sqrt{\calR_c+3\sG^4}}{\sG},
\end{equation}
one finds
\begin{equation}
    \beta_{{\rm G}^4}=\erfc\left(\frac{|y_c|}{\sqrt{2}}\right), \qquad \Rightarrow \,\, |y_c|=\sqrt{2} \inverfc\left(\beta_{{\rm G}^4}\right),
\end{equation}
which we can rearrange (using $\langle\calRG^8\rangle=7!!\sG^8$) to find the variance is 
\begin{equation}
    A_{G^4}=96A_{\rm G}^4=\frac{96\,\calR_c^2}{\left(4\,\inverfc^4\left(\beta_{{\rm G}^4}\right)-3\right)^2}.
\end{equation}

\printbibliography

\end{document}